\documentclass{article}
\usepackage{psfig}
\usepackage{emulateapj}

\lefthead{Nevalainen J., Markevitch M. and  Forman W.}
\righthead{X-ray total mass estimate for A3571}

\begin{document}

\submitted{Accepted to ApJ}

\title{X-ray total mass estimate for the nearby relaxed cluster A3571}

\author{J. Nevalainen\altaffilmark{1}, M. Markevitch\altaffilmark{2} and W. Forman}
\affil{Harvard Smithsonian Center for Astrophysics, Cambridge, USA}

\altaffiltext{1}{Observatory, University of Helsinki, Finland}
\altaffiltext{2}{Space Research Institute, Russian Acad. of Sci.}

\begin{abstract}
We constrain the total mass distribution in the cluster A3571, combining spatially resolved ASCA temperature data with ROSAT imaging data with the assumption that the cluster is in hydrostatic equilibrium.
The total mass within $r_{500}$ (1.7 $h_{50}^{-1}$ Mpc) is
$M_{500}$ = $7.8^{+1.4}_{-2.2} \times 10^{14} \, h_{50}^{-1} \, M_{\odot}$ 
at 90\% confidence, 
1.1 times smaller than the isothermal estimate.
The Navarro, Frenk \& White ``universal profile'' is a good description of the dark matter density distribution in A3571. The gas density profile is shallower than the dark matter profile, scaling as 
$r^{-2.1}$ at large radii, leading to a monotonically increasing gas mass fraction with radius. Within $r_{500}$ the gas mass fraction reaches a value of 
$f_{gas} = 0.19^{+0.06}_{-0.03} \ h_{50}^{-3/2}$
(90\% confidence errors).
Assuming that this value of $f_{gas}$ is a lower limit for the the universal value of the baryon fraction, we estimate the 90\% confidence upper limit of the cosmological matter density to be 
$\Omega_{m} < 0.4$.
\end{abstract}


\section{Introduction}
Measuring cluster masses has significant implications for cosmology. Assuming that the cluster mass content represents that of the Universe, measured total and baryonic mass distributions in a cluster, 
combined with the Big Bang nucleosynthesis calculations and observed light element abundances, can be used to constrain the cosmological density parameter (White et al. 1993). A measured cluster mass 
function could be used to constrain cosmological parameters via the Press-Schechter formalism. However, the mass cannot be observed directly since most of the cluster mass is dark matter, and one must rely
on observed quantities like gas temperature or galaxy velocities and assume the state of the cluster to derive the mass.

In this paper, we estimate the total mass for the A3571 cluster under the assumptions of hydrostatic equilibrium and thermal pressure support. Until recently, most hydrostatic X-ray mass estimates have been
made assuming that the gas is isothermal at the average broad beam temperature. However, the total mass within large radii is only as accurate as the local temperature at that radius. ASCA observations 
provide spatially resolved temperature data for hot clusters and yield their 2D temperature structure. A large number of ASCA clusters shows that the temperature declines with increasing radius 
(\cite{tmaps}), in qualitative accordance with hydrodynamic cluster simulations (e.g. \cite{evr96}, \cite{bn}, \cite{burns}). This implies that the total mass within small radii is greater, while at large 
cluster radii it is smaller than that derived assuming isothermality, which has also been observed in A2256 (\cite{A2256}, A2029 (\cite{sar2029}), A496  and A2199 (\cite{A496_A2199}) and A401 (\cite{A401}).
Consequently, the gas mass fraction within a large radius where the cluster may be a fair sample of the universe, is larger than that derived assuming isothermality which further aggravates the ``baryon 
catastrophe'' (e.g. \cite{wnef}, \cite{wf}, \cite{etto99}, \cite{mohr}). 

A3571 (z = 0.040) is suitable for measuring the dark and total mass distributions, since it is bright and hot ($\sim$ 7 keV) allowing accurate temperature determinations with ASCA. Indeed, the A3571 
temperature profile used in this work (see \cite{tmaps}) is among the most accurate for all hot clusters. The three ASCA pointings cover the cluster to $r_{500}$ (the radius where the mean interior density 
equals 500 times the critical density, approximately the radius inside which hydrostatic equilibrium holds, according to simulations of \cite{evr96}). A3571 has a cooling flow (\cite{peres}), but it is weak 
enough not to introduce  large uncertainties in the temperature determination.

We use $H_{0} \equiv 50 \, h_{50} \, {\rm km} \, {\rm s}^{-1} \, {\rm Mpc}^{-1}$, $\Omega = 1$ and report 90\% confidence intervals  throughout the paper.

\section{ROSAT ANALYSIS}
\label{ROSAT}
\begin{figure*}
\psfig{figure=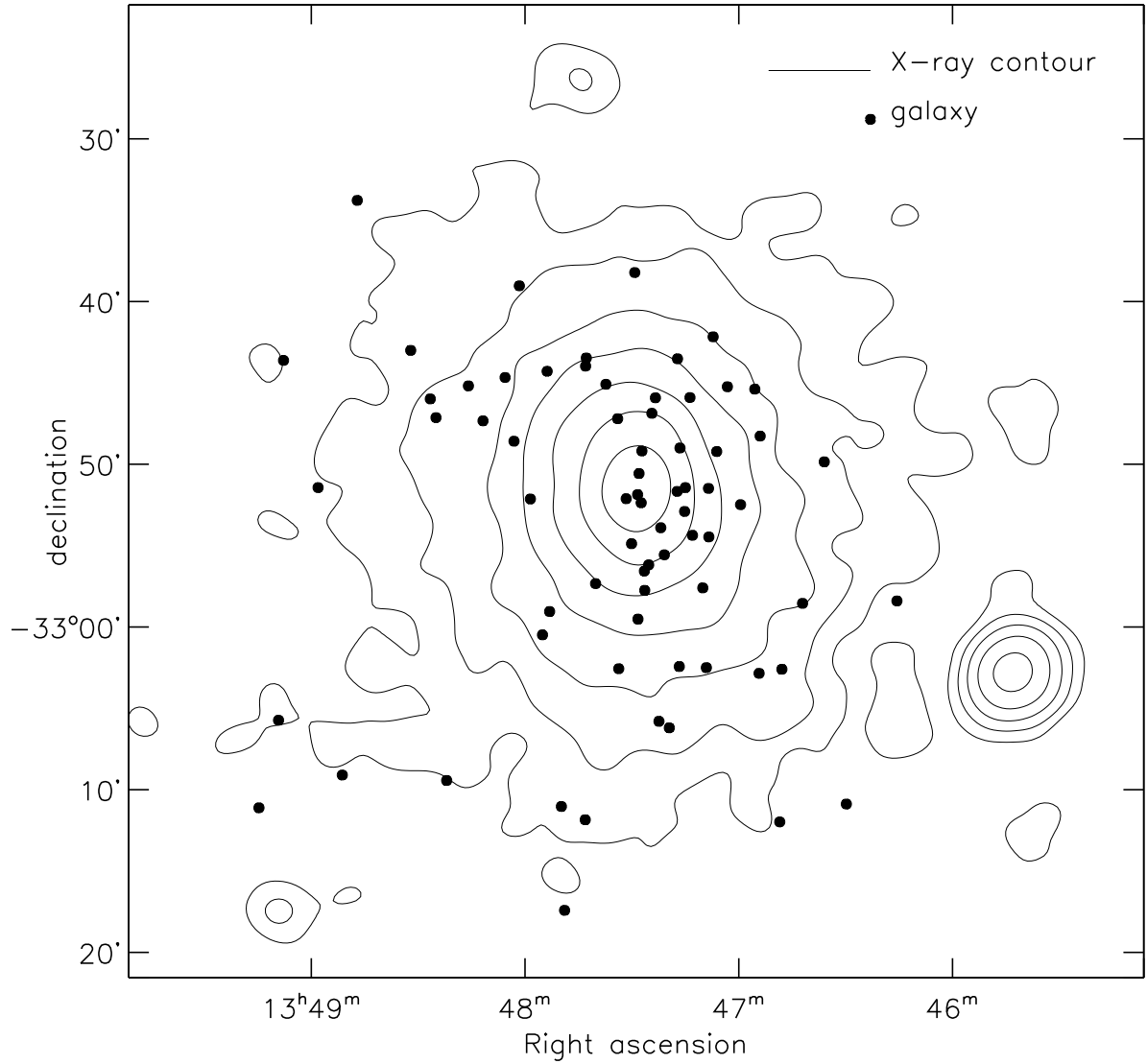,width=19cm,angle=0} 
\vspace{1cm}
\caption{ROSAT PSPC contour map of the surface brightness of A3571 in 0.73-2.04 keV energy range, smoothed by a Gaussian with $\sigma = 1'$.  The contour level values are 0.0005, 0.001, 0.002, 0.004, 
0.008, 0.016 and 0.032 counts s$^{-1}$ arcmin$^{-2}$. Galaxies from the NASA Extragalactic Database are plotted as filled circles. \label{fig1}}
\end{figure*}

We processed the ROSAT data, consisting of a PSPC pointing rp800287, using Snowden's Soft X-Ray Background programs (\cite{sno}), which reduced the total exposure by 25\% to 4.5 ks. The spatial analysis was 
restricted to the energy band of 0.73 - 2.04 keV (Snowden's bands R6-R7) to improve sensitivity over the X-ray background. The surface brightness contour map (smoothed by a Gaussian with $\sigma$ = 1$'$) is 
shown in Figure \ref{fig1}a. The data show no obvious substructures and no deviations from azimuthal symmetry, except for a slight ellipticity. Fabricant et al. (1984) showed that for A2256, whose X-ray 
brightness distribution is more elliptical than the one of A3571, the true elliptical total mass is very close to the hydrostatic total mass derived assuming spherical symmetry. Furthermore, Vikhlinin et al.
(1999) divided the PSPC data of A3571 into several sectors, fitted the brightness with an azimuthally symmetric model and found that the azimuthal variation of the gas density gradient (due to ellipticity) 
was 6\% of the global value, which would indicate a similar error in the total mass, which is negligible compared to total mass errors obtained with spherical model. Therefore, in the following analysis we 
assume the cluster to be azimuthally symmetric.

\begin{figure*}
\psfig{figure=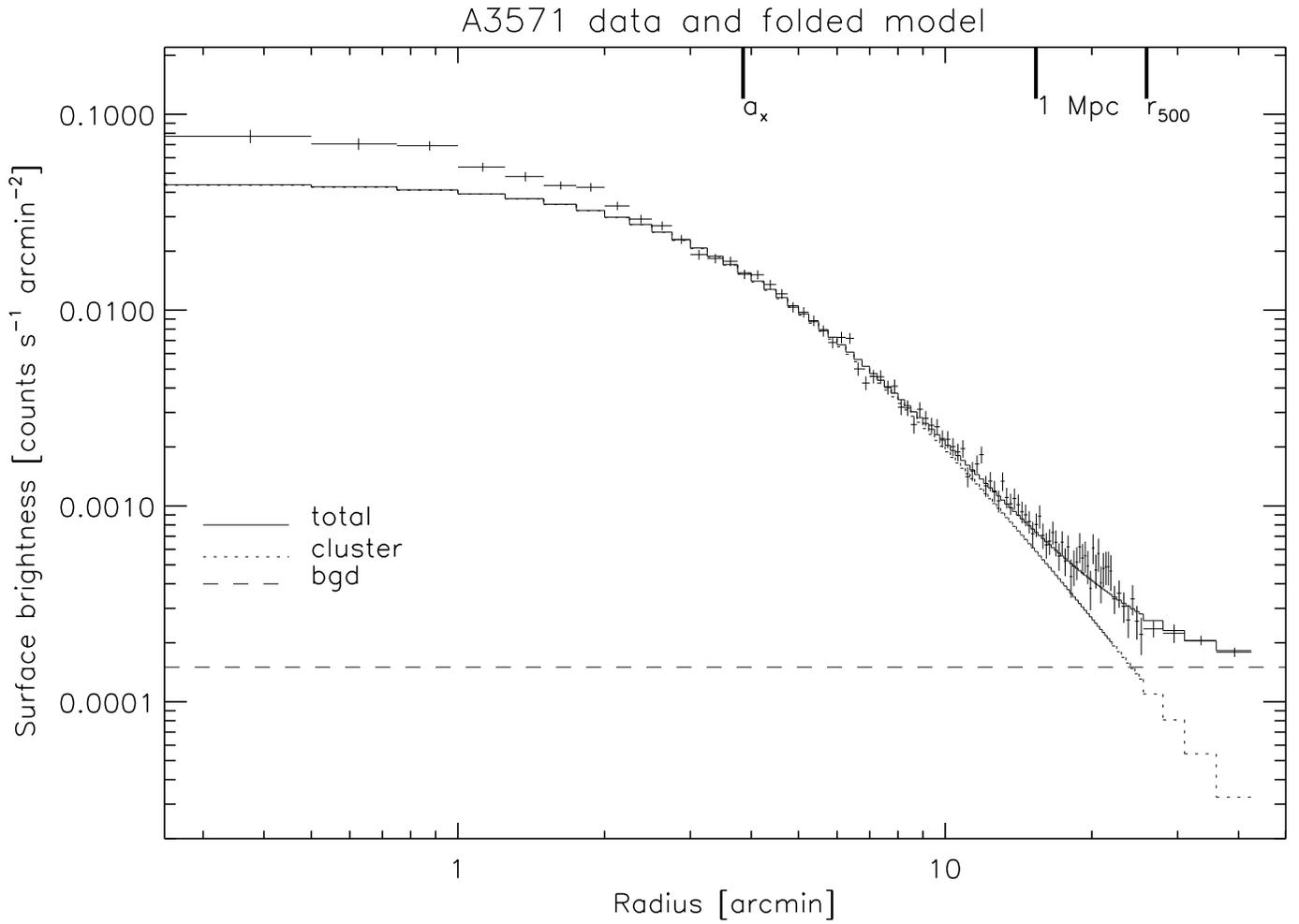,width=19cm,angle=0} 
\caption{ROSAT PSPC radial surface brightness profile in 0.73-2.04 keV range together with the PSF-convolved best-fit $\beta$ model.
The data with $r < 3'$ are not included in the fit, due to central cooling flow. \label{surfbright_plot}}
\end{figure*}

We excluded point sources and generated a radial surface brightness profile in concentric annuli of width ranging from $15''$ at the center to $7'$ at a radial distance of $40'$. In the radial range of ASCA 
pointings ($r < 35'$), we included only the ROSAT data from the sky areas covered by ASCA. A cooling flow with a mass flow rate of 40 - 130 M$_{\odot}$/yr and a cooling radius of 1.2 - 2.2 arcmin has been 
detected in the center of A3571 by Peres et al. (1998). Our data show a significant central brightness excess over the $\beta$-model (Figure \ref{surfbright_plot}), in agreement with the reported cooling 
flow. Therefore  we excluded the data within the central $r < 3'$ from the fit.

We fitted the observed profile with a $\beta$- model  
\begin{equation}
I(b) = I_{0}\left(1 + \left(\frac{b}{a_{x}}\right)^{2}\right)^{(-3 \beta + \frac{1}{2})} + CXRB
\label{cav} 
\end{equation}
(\cite{cava}), where $b$ is the projected radius. We fixed the cosmic X-ray background (CXRB) to $1.5 \times 10^{-4}$ counts s$^{-1}$ arcmin$^{-2}$ found from the outer part of the image, and included 5\% of
the background value as a systematic error, due to variation on the sky. We used XSPEC to convolve the surface brightness model through a spatial response matrix (constructed from the ROSAT PSF at 1 keV, for
an azimuthally symmetric source centered on-axis) and to compare the convolved profile with the data. We find an acceptable fit in the radial range 3$'$-43$'$ (see Figure \ref{surfbright_plot} and 
Table \ref{tab1}), with best fit parameters $a_{x} = 3.85 \pm 0.35$ arcmin (= $ 310 \pm 30 \ h_{50}^{-1} kpc $), $\beta = 0.68 \pm 0.03 $ with $\chi^2$ = 74.1 for 87 degrees of freedom. Our values of 
$a_{x}$ and $\beta$ are consistent with another study of the ROSAT PSPC data of A3571 (Vikhlinin et al. 1999) who also excluded the cooling flow area from the fit. In yet another study of ROSAT PSPC data of 
A3571 (Mohr et al. 1999) inconsistently smaller values were found for $a_{x}$ and $\beta$. This is a consequence of including the data of the cooling flow into the profile fit in that work. 

If we assume that the intracluster gas is isothermal and spherically symmetric, the best-fit parameters $a_{x}$ and $\beta$ determine the shape of the gas density profile as:
\begin{equation}
\rho_{gas}(r) = \rho_{gas}(0)\left(1 + \left(\frac{r}{a_{x}}\right)^{2}\right)^{ -\frac{3}{2} \beta}
\end{equation}
The observed temperature variation in A3571 from 7 keV to 4 keV with radius will introduce at most a 2\% effect on the gas mass (e.g. Mohr et al. 1999), which is negligible compared to other components in 
our error budget.

We obtained the normalization of the gas density profile (as in \cite{vikh}) $\rho_{gas}(0) = 1.5 \times 10^{14}$M$_{\odot}$ Mpc$^{-3}$, or 1.0 $\times 10^{-26}$ g cm$^{-3}$, by equating the emission measure 
calculated from the above equation, with an observed value of 8.1 $\times 10^{67}$ cm$^{-3}$ inside a cylinder with $r = 0.1 - 2$ $h_{50}^{-1}$ Mpc radius, centered at the cluster brightness peak
($r = 0.1$ Mpc encompasses the cooling flow excluded from all our analyses).

\section{TEMPERATURE DATA}
We used the temperature profile data presented in Markevitch et al. \ (1998), who combined three ASCA pointings to derive the emission weighted, cooling flow - corrected temperature kT = $ 6.9 \pm 0.2 $ keV. 
Outside the cooling radius, the temperature values were measured in radial bins 2$'$-6$'$-13$'$-22$'$-35$'$ (0.13-0.39-0.85-1.43-2.28 $h_{50}^{-1}$ Mpc).
The central bin that is affected by a significant cooling flow component is not used in the analysis below. The temperature errors were determined by generating Monte - Carlo data sets which properly  
account for the statistical and systematic uncertainties (including those of the PSF, effective area and background). As in most nearby clusters, the ASCA data reveal a temperature decline with radius. The 
ROSAT PSPC data on A3571 in the 0.2--2 keV band were also analyzed by Irwin et al. (1999), who derive a temperature profile consistent with a constant up to 20$'$. However, those authors did not include the 
PSPC calibration uncertainties that dominate the ROSAT temperature errors for hot clusters such as A3571 (see e.g. \cite{mv1}); inclusion of these uncertaintines should make their results consistent with the
ASCA profile. The ASCA temperature profile for A3571 is similar to profiles of a large sample of nearby ASCA clusters (\cite{tmaps}), when scaled to physically meaningful units of the radii of fixed 
overdensity. Therefore it appears unlikely that the observed decline is due to an unknown instrumental effect. A more detailed discussion about the validity of the ASCA spatially resolved temperature 
data can be found in Nevalainen et al.\ (1999a). Hydrodynamic simulations predict a qualitatively similar radial temperature behavior in relaxed clusters (e.g., \cite{evr96}; Eke, Navarro, \& Frenk 1997; 
\cite{bn}), although there are differences in detail between the simulations and observations as well as between the different simulation techniques (e.g., Frenk et al.\ 1999).

\section{VALIDITY OF THE HYDROSTATIC EQUILIBRIUM}
Quintana \& de Souza (1993) report preliminary results of an optical study of the galaxies in A3571. They find a suggestion that the galaxy distribution in A3571 is irregular and forms several velocity 
subgroups, but they did not perform  quantitative statistical analyses of the galaxy distribution, due to the small number of observed galaxies (see Figure 1 for the distribution of A3571 member galaxies 
from the NASA Extragalactic Database). The central giant galaxy MCG05-33-002 has an extensive optical halo with dimensions of 0.2 $\times$ 0.6 $h_{50}^{-1} $Mpc, elongated along the major axis of the core 
region of this galaxy (\cite{kemp}). The galaxy distribution of A3571 is also aligned in the same direction (\cite{kemp}). As discussed by Quintana \& de Souza (1993), the optical data suggests that the cD 
galaxy formed during the original collapse of the central part of the cluster and that the cluster may not yet be virialized. However, as Quintana \& de Souza (1993) state, their galaxy distribution results
are only tentative. Furthermore, galaxies are not the best measure of the relaxation since clusters form within intersecting filaments in larger scale and superpositions can give the appearance of the 
asymmetries, substructure, and superposed groups.

In X-rays, the ROSAT PSPC data of A3571 (Figure 1) show that the gas is azimuthally symmetric (except for a slight ellipticity, see Section \ref{ROSAT}) and that there is no substructure and no correlation 
between the galaxy and gas distributions at large radii. Furthermore, the ASCA gas temperature map of A3571 (\cite{tmaps}) shows no asymmetric variation that would indicate dynamic activity. 

Neumann \& Arnaud (1999) found evidence in their ROSAT cluster sample that cooling flows are a recurrent phenomena that may be turned off by mergers, in accordance with a hierarchical clustering scenario
(Fabian et al. 1994). Since A3571 has a considerable cooling flow (Peres et al. 1998), any merger must have been either not very strong or sufficiently in the past for the gas to reestablish equilibrium and
a cooling flow.

A3571 is a member of the Shapley supercluster (Raychaudhury et al. 1991) and therefore likely to have more frequent mergers and may not be typical of more isolated clusters. However, all the X-ray 
evidence consistently argues against any significant ongoing merger in A3571. Since the optical evidence does not contradict significantly the X-ray evidence for non-merger, we assume that the hydrostatic 
equilibrium is valid in A3571.

\section{MASS FITTING}
\subsection{Method}

For the details of the mass calculation, we refer to our similar analysis of cluster A401 (\cite{A401}). Briefly, we model the dark matter density with a constant core model 
\begin{equation}
\rho_{dark} \propto \left(1 + \frac{r^2}{a_d^2}\right)^{- \alpha/2},
\label{coremodel}
\end{equation}
and with the central cusp profile:
\begin{equation}
\rho_{dark} \propto \left(\frac{r}{a_{d}}\right)^{- \eta} \left(1 + \frac{r}{a_{d}}\right)^{\eta - \alpha}.
\label{cuspmodel}
\end{equation}
We fix $\eta$ = 1 in the cusp models, as suggested by numerical simulations (\cite{nfw}, hereafter NFW), but vary the other parameters. We solve the hydrostatic equilibrium equation 
\begin{equation}
M_{tot}(\le r) = 3.70 \times10^{13} M_{\odot} {T(r) \over {\rm keV}} {r \over {\rm Mpc}} \left( - {{d \ln{\rho_{gas}}} \over {d \ln{r}}} - {{d \ln{T}} \over {d \ln{r}}} \right),
\label{hydreq}
\end{equation}
(e.g. Sarazin 1988, using $\mu = 0.60$), for temperature, in terms of dark matter and gas density profile parameters. We fix the gas density to that found from the ROSAT data above, calculate the 
3 - dimensional temperature profile model corresponding to given dark matter parameters, project it on the ASCA annuli, compare these values to the observed temperatures and iteratively determine the dark 
matter distribution parameters. To propagate the errors of the temperature profile data to our mass values, we repeat the procedure for a large number of Monte - Carlo temperature profiles with added random 
errors. We reject unphysical models that give infinite temperatures at large radii, and those models that are convectively unstable (that is, correspond to polytropic index $ > \frac{5}{3} $ in the radial 
range of the temperature data, outside the cooling flow region  $r$ = 3$'$-35$'$). From the distribution of the acceptable Monte - Carlo models, we determine the 1 $\sigma$ confidence intervals of the mass 
values as a function of radius. We convert these values to 90\% confidence values, assuming a Gaussian probability distribution. We cannot constrain all dark matter model parameters independently due 
to the limited accuracy of the temperature data. However, the models with steeper dark matter density slopes (higher $\alpha$) require larger dark matter core radii (higher $a_{d}$) to produce similar shapes 
of temperature profile and due to this correlation the corresponding mass values vary within a relatively narrow range. We also propagate the estimate of the uncertainty of the local gas density gradient to 
the total mass values, as in Nevalainen et al. (1999a).

\begin{figure*}
\psfig{figure=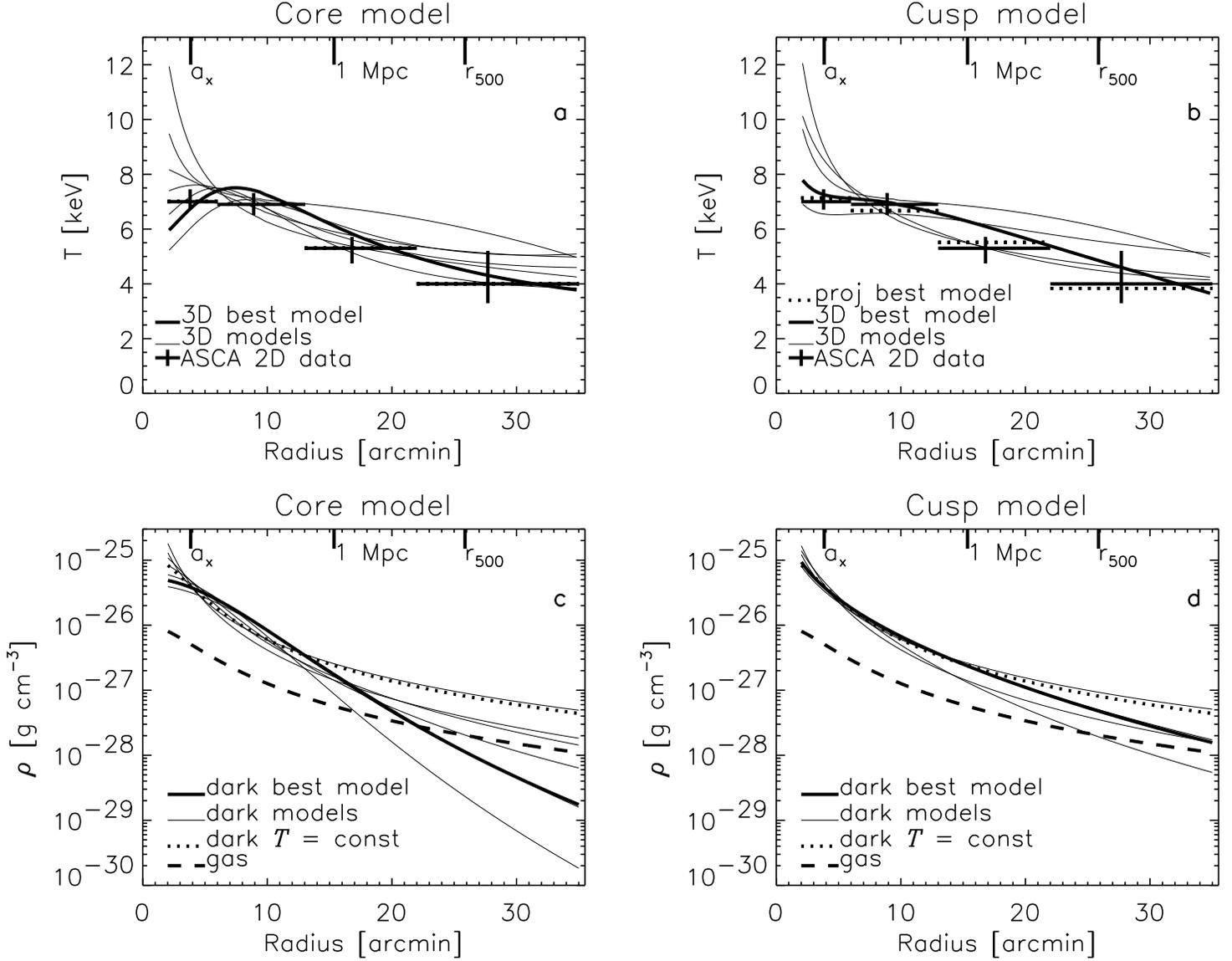,width=19cm,angle=0} 
\vspace{1cm}
\caption{In (a) and (b), crosses show projected ASCA temperatures with 1 $\sigma$ errors. Thin solid lines show a representative set of temperature models of the constant core form (a) and the cusp form
(b), before projection, allowed by the convective stability constraint (see text). Thick solid line shows the best fit model (before projection). In the case of cusp model, the thick dotted line shows
the values of this best model projected to the 2D ASCA bins, which are compared with the ASCA data. In (c) and (d), the corresponding density models are plotted, together with the gas density model. For 
comparison, values assuming isothermality (kT = 6.9 keV) are also shown. \label{temperatures_plot}}
\end{figure*}

In Figure \ref{temperatures_plot} we show representative density profiles of forms (\ref{coremodel}) and (\ref{cuspmodel}), and the corresponding model temperature profiles. Both functional forms give 
acceptable fits to the data and yield masses consistent within 90\% confidence errors. Our final 90\% confidence intervals of total mass, at each radius, include the 90\% confidence intervals of both models,
and the average of the two models is used as the best value. As can be seen in Figures \ref{temperatures_plot}c and \ref{temperatures_plot}d, in the radial range $3'-15'$, the Monte-Carlo densities are lower
than the best fit values. This asymmetry is due to the fact that the polytropic index criterion effectively rejects the most massive models, as was also found for  A401 (Nevalainen et al. 1999a).  

\subsection{Results}
\begin{figure*}
\psfig{figure=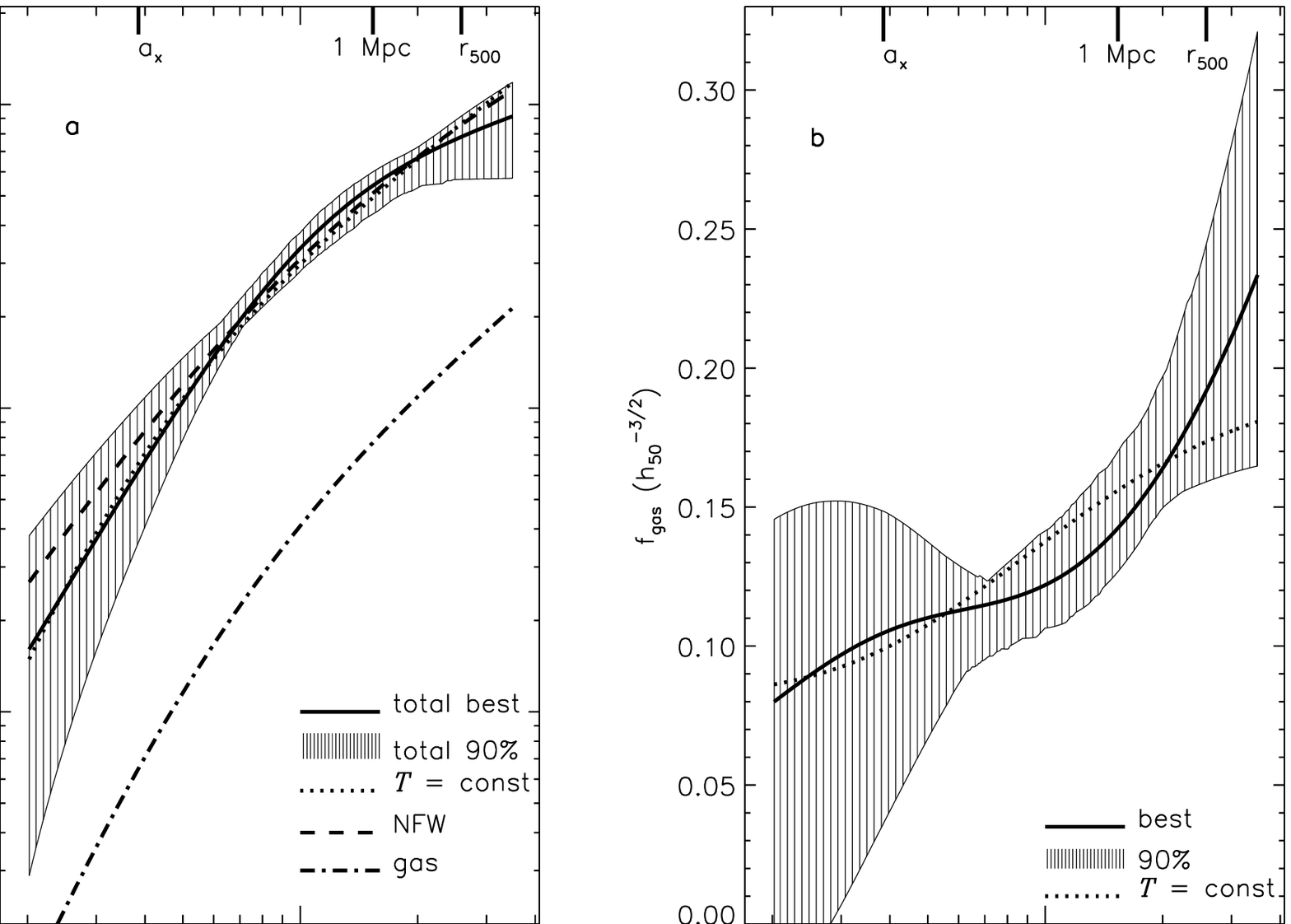,width=16cm,angle=0} 
\vspace{1cm}
\figcaption{(a) shows the enclosed mass profile (solid line) with 90\% confidence errors (shaded area), obtained combining the core and the cusp model fit results. 
Also the model assuming  isothermality (dotted line), together with the gas mass profile (dash-dot) are shown. The best fit ``universal'' dark matter profile, suggested by NFW simulations, is plotted as 
dashed line. In (b), the gas mass fraction with errors is shown. Masses are evaluated using $H_{0} = 50$ km$^{-1}$ Mpc$^{-1}$ (total mass scales as $H^{-1}$ and gas mass as  $H^{-5/2}$).\label{masses_plot}}
\end{figure*}

The final mass profile is shown in Fig. \ref{masses_plot}. The overdensity, or the mean interior density in units of the critical density, calculated from our best fit models, is 240 at $r = 35'$, the 
largest radius covered by the ASCA data. Simulations (e.g. \cite{evr96}) suggest that within $r_{500}$, where the overdensity is 500,  hydrostatic equilibrium is valid. For A3571,
\begin{equation}
r_{500} = 25.9' = 1.7 \ h_{50}^{-1} \ {\rm Mpc}
\label{r500eq}
\end{equation}

and the mass within this radius 

\begin{equation}
M_{tot}(\le r_{500}) = 7.8^{+1.4}_{-2.2} \times 10^{14} h_{50}^{-1} \ M_{\odot}.
\label{M500_eq}
\end{equation}
The mass values within several interesting radii are given in Table 1.

At large radii our mass errors are quite large because they cover the values allowed by two different models, and include the uncertainty of $\beta$. Therefore the isothermal mass is consistent with our 
results, but compared to our best values, the isothermal ones are greater by factors of 1.1 and 1.3 at radii of $r_{500}$ and 35$'$ (see Figure \ref{masses_plot}). This difference is a natural consequence of 
the real temperatures being lower than the average temperature at large radii, similarly with other clusters with measured temperature profiles (\cite{A2256}, \cite{A401}, \cite{A496_A2199}). However, at 
small radii ($ r = a_{x}$), differently from the other above clusters, the isothermal mass in A3571 is about equal to the value obtained with the observed temperature profile, due to the nearly constant 
temperature up to $13'$ in A3571. 

The deprojection method, with the isothermal assumption, gives a total mass value of $6.22 \times 10^{14} h_{50}^{-1} \ M_{\odot}$ inside a radius of 0.91 $h_{50}^{-1}$ Mpc (\cite{etto97}), whereas our value
at that radius is significantly smaller, $4.9^{+0.6}_{-0.9} \times 10^{14} h_{50}^{-1} \ M_{\odot}$. This behaviour is similar to that found for A401 (\cite{A401}).

The frequently used mass - temperature scaling law obtained in cosmological simulations (\cite{evr96}) predicts that A3571, which has kT = $ 6.9 \pm 0.2$ keV (Markevitch et al. 1998), will have 
$r_{500} = 2.1 \pm 0.14 \ h_{50}^{-1} \ {\rm Mpc} $ and 
$M_{tot}(\le r_{500}) = 1.3 \pm 0.19 \times 10^{15} h_{50}^{-1} \ M_{\odot}$,
whereas our measured values above are smaller by factors of 1.2 and 1.6. Similar behaviour has been found in several other hot clusters with temperature profiles measured with ASCA, i.e. A2256 
(\cite{A2256}), A401 (\cite{A401}), A496 and A2199 (\cite{A496_A2199}). Also temperature profiles of cooler groups NGC5044 and HCG62 and a galaxy NGC507, measured with ROSAT, give similar results 
(Nevalainen et al. 1999b). These comparisons suggest that the above simulations produce too small temperature for a given mass.

The virial theorem analysis of the galaxy velocity distribution in A3571 (Girardi et al. 1998) gives $R_{vir} = 4.18 h_{50}^{-1}$ Mpc and $M_{vir} = 1.63^{+0.79}_{-0.72} \times 10^{15} h_{50}^{-1} M_{\odot}$,
whereas our values extrapolated to this radius are $1.2 \pm 0.7 \times 10^{15} h_{50}^{-1} M_{\odot}$, consistent within the large errors.

\section{DISCUSSION}
\subsection{NFW profile}
With $\eta = 1$ and $\alpha = 3$ the cusp model (Eq. 4) corresponds to  the NFW ``universal'' mass profile. For A3571 there is a significant detection of a cooling flow in the center and the polytropic 
$\gamma \le \frac{5}{3}$ constraint is applicable only beyond the cooling flow region. At those radii the temperature gradient of the cusp model is not strong, and the model is convectively stable.  
This model also gives an acceptable fit to the A3571 data, and is consistent with our mass profile within the errors (see Figure \ref{masses_plot}). Using the best fit NFW profile for A3571, we obtain the 
concentration parameter $c \equiv r_{200}/a_{d} = 5.3$  and $M_{200} = 1.3 \times 10^{15} h_{50}^{-1} \ M_{\odot}$. These values are consistent with the NFW simulations in 
SCDM and CDM$\Lambda$ cosmological models.  

In the hydrostatic equilibrium scheme, since the observed gas density and temperature profiles are similar in different clusters, when scaled by their estimated virial radii (\cite{vikh} and \cite{tmaps}, 
respectively), a similar total mass distribution is implied, and that is what we observe. The shapes of the total mass profiles at large radii in other hot clusters with measured ASCA temperature profiles 
for A2256 (\cite{A2256}), A2029 (\cite{sar2029}), A496 and A2199 (\cite{A496_A2199}) and A401 (\cite{A401}) are also consistent with the NFW model. NFW profile also describes well the mass profiles of cool 
groups NGC5044 and HCG62 and a galaxy NGC507 that are derived using ROSAT PSPC temperature profiles (David et al. 1994, Ponman \& Bertram 1993,  Kim \& Fabbiano 1995, respectively, see the discussion of NFW
models for these objects in Nevalainen et al. 1999b). These consistencies suggest that the NFW profile may indeed be universal.  

\subsection{Gas mass fraction}
The dark matter density in the best fit model falls as r$^{-4}$ at large radii, whereas the gas density falls as r$^{-2}$. This causes the gas mass fraction to increase rapidly at large radii, to a value of
\begin{equation}
f_{gas}(\le r_{500}) = 0.19^{+0.06}_{-0.03} \, h_{50}^{-3/2}.
\label{fgaseq}
\end{equation}
(see Figure \ref{masses_plot}). This value is consistent with those for A2256 (\cite{A2256}), A401 (\cite{A401}), A496 and A2199 (\cite{A496_A2199}) obtained using ASCA temperature profiles implying that the
gas and dark matter distributions are similar in different clusters. Our value is also consistent with results for samples of clusters analyzed assuming isothermality:
$f_{gas}(\le r_{500}) = 0.168^{+0.065}_{-0.056} \, h_{50}^{-3/2}$ (\cite{etto99}) and 
$f_{gas}(\le r_{500}) = 0.212 \pm 0.006 \, h_{50}^{-3/2}$ (for clusters with $kT > 5$ keV \cite{mohr}).
At larger radii the temperature profile analysis would probably yield still higher values compared to the isothermal ones.

Following the method of White et al. (1993; see also \cite{A401} for its application to A401), using the $f_{gas}$ value above,  we compute an estimate for the cosmological density parameter 
$\Omega_{m} = <\rho>/\rho_{crit}$ as
\begin{equation}
\Omega_{m} = \Upsilon \, \Omega_{b}  {\left(f_{gas} + {M_{gal} \over M_{tot}}\right)}^{-1},
\label{omegaeq}
\end{equation}
at $r_{500}$ for A3571, 
where $\Upsilon$ is the local baryon diminution, for which we use a value 0.90 as suggested by simulations (Frenk et al. 1999).
Due to the lack of a reliable estimate of the total galaxy mass in A3571 within $r_{500}$ we compute only the upper limit for $\Omega_{m}$ using $M_{gal} = 0$.
We take $\Omega_{b} = 0.076 \pm 0.007$ (\cite{burl}). 
Using a reasonable lower limit for the Hubble constant $H_{0} > $ 60  km s$^{-1}$ Mpc$^{-1}$ (e.g. Nevalainen \& Roos 1998) we obtain
$\Omega_{m}  < 0.4$, consistent with several independendent current $\Omega_{m}$ estimates (Freedman 1999, Roos \& Harun-or-Rashid 1999). 

\section{CONCLUSIONS}

We have constrained the dark matter distribution in A3571, using accurate ASCA gas temperature data. The dark matter density in the best fit model scales as r$^{-4}$ at large radii, and the NWF profile also 
provides a good description of the dark matter density distribution. The total mass within $r_{500}$ (1.7 $h_{50}^{-1}$ Mpc) is $7.8^{+1.4}_{-2.2} \times 10^{14} \, h_{50}^{-1} \, M_{\odot}$ at 90\% 
confidence, or  1.1 times  smaller than the isothermal value, or 1.6 times smaller than that predicted by the scaling law based on simulations (Evrard et al. 1996), which is qualitatively similar to the 
results for other clusters with accurate temperature profiles. The gas density profile in A3571, proportional to  $r^{-2.1}$ at large radii, is shallower than that of the dark matter. Hence the gas mass 
fraction increases with radius, with
$f_{gas}(r_{500}) = 0.19^{+0.06}_{-0.03} \, h_{50}^{-3/2}$ (90 \% errors)
at $r_{500}$, consistent with results for A2256, A401, A496, and A2199.
Assuming that this is a lower limit of the primordial baryonic fraction, we obtain $\Omega_{m} < 0.4$ at 90\% confidence. However, $f_{gas}$ is still strongly increasing at $r_{500}$, so that we obviously 
have not reached the universal value of the baryon fraction, which would make $\Omega_{m}$ even smaller.
	
\acknowledgments
JN thanks Harvard Smithsonian Center for Astrophysics for the hospitality. JN thanks the Smithsonian Institute for a Predoctoral Fellowship, and the Finnish Academy for a supplementary grant. We are indebted
to Dr A.Vikhlinin for several helpful discussions. We thank Prof. M.Roos for his help. WF and MM acknowledge support from NASA contract NAS8-39073. This research has made use of the NASA/IPAC Extragalactic 
Database (NED) which is operated by the Jet Propulsion Laboratory, California Institute of Technology, under contract with the National Aeronautics and Space Administration. We thank the referee for a 
careful report on our paper.

\clearpage

\begin{deluxetable}{lr}
\tablecaption{Mass of A3571 \tablenotemark{*} \label{tab1}}
\tablewidth{0pt}
\tablehead{
\colhead{Parameter} & \colhead{Value}}
\startdata
$a_x$ \tablenotemark{a} \ [arcmin]                                                              &  $3.85 \pm 0.35$                     \\
$a_x$ \tablenotemark{a} \ [$h_{50}^{-1} \ {\rm kpc}$]                                           &  $310  \pm 30 $                      \\
$\beta$ \tablenotemark{a}                                                                     &  $0.68 \pm 0.03$                     \\
                                                                                              &                                      \\
$\rho_{gas}$(0) [$10^{-26} \ h_{50}^{1/2} \ {\rm g} \ {\rm cm}^{-3}$]                         &  1.0                                 \\
                                                                                              &                                      \\
M($< 3.85' = a_{x}$)           [$M_{\odot}$]                                                          &  $6.2^{+4.1}_{-2.6} \times 10^{13}$  \\
M($< 15.2' = $ 1 Mpc)           [$M_{\odot}$]                                                          &  $5.4^{+0.6}_{-1.1} \times 10^{14}$  \\
M($< 25.9' = r_{500}$) [$M_{\odot}$]                                                          &  $7.8^{+1.4}_{-2.2} \times 10^{14}$  \\
M($< 35'$)   [$M_{\odot}$]                                                          &  $9.1^{+2.8}_{-4.4} \times 10^{14}$  \\
$f_{gas}$ ($< 25.9' = r_{500}$) $\times$ $h_{50}^{3/2}$                                       &  $0.19^{+0.06}_{-0.03}$               \\
\tablenotetext{*} {using $H_{0} \equiv 50 \ h_{50} \ km \ s^{-1} \ Mpc^{-1}$. All errors are 90\% confidence errors.}\\
\tablenotetext{a}{From ROSAT PSPC data excluding the central $r \le 3'$ cooling flow region.}\\
\enddata
\end{deluxetable}

\clearpage

\clearpage

\clearpage

\clearpage

\begin{thebibliography}{}


\bibitem[Bahcall \& Sarazin 1977]{bahc} Bahcall, J.N. and Sarazin, C.L., 1977, ApJL, 213:99-103.  
\bibitem[Bryan \& Norman 1997]{bn} Bryan, G.L.,  Norman, M.L., {\it Computational Astrophysics} Proc. 12th Kingston Conference, Halifax, Oct. 1996, ed. D. Clarke \& M. West (PASP), also astro-ph/9710186 .
\bibitem[Burles et al. 1998]{burl} Burles, S., Tytler, D., 1998, The Proceedings of the Second Oak Ridge Symposium on Atomic \& Nuclear Astrophysics, (Oak Ridge, TN, December 2-6, 1997), ed. A. Mezzacappa
(Institute of Physics, Bristol), p.113, also astro-ph/9803071.
\bibitem[Burns et al. 1999]{burns} Burns, J., Bryan, G. \& Norman, M.L., 1999, Proc, Texas Symp. 
\bibitem[Cavaliere \& Fusco-Femiano 1976]{cava} Cavaliere and Fusco-Femiano, 1976,AA, 49,137.
\bibitem[Churazov et al. 1996]{chur} Churazov, E., Gilfanov, M., Forman, W. \& Jones, C.
\bibitem[David et al. 1994]{dav} David, L.P., Jones, C., Forman, W. \& Daines, S., 1994, ApJ, 428, 544.
\bibitem[Donnelly et al. 1999]{donn1} Donnelly, R.H., Markevitch, M., Forman, W., Jones, C., Churazov, E. \& Gilfanov, M., 1999, ApJ, 513, 690.
\bibitem[Donnelly et al. 1998]{donn2} Donnelly, R.H., Markevitch, M., Forman, W., Jones, C., David, L.P., Churazov, E. \& Gilfanov, M., 1998, ApJ, 500, 138.
\bibitem[Eke et al. 1998]{eke} Eke, V. R., Navarro, J. F., \& Frenk, C. S. 1998, ApJ, 503, 569. 
\bibitem[Evrard et al. 1996]{evr96} Evrard, A., Metzler, C. and Navarro, J., 1996, ApJ 496, 494.
\bibitem[Ettori \& Fabian 1999]{etto99} Ettori, S. \&  Fabian, A.C., 1999, MNRAS, 305, 834.
\bibitem[Ettori et al. 1997]{etto97} Ettori, S., Fabian, A.C. \& White, D.A., 1997, MNRAS 289, 787.
\bibitem[Fabian et al. 1994]{fab} Fabian, A. C., Crawford, C. S., Edge, A. C. \& Mushotzky, R. F., 1994, MNRAS, 267, 779.
\bibitem[Freedman 1999]{fre} Freedman, W, 1999, {\it Determination of Cosmological Parameters}, World Scientific Press, submitted, astro-ph/9905222.
\bibitem[Frenk et al. 1999]{fre99} Frenk, C.S.,  et al. 1999 ApJ, 519, 518.
\bibitem[Ikebe et al. 1997]{ike} Ikebe, Y., Makishima, K., Ezawa, H., Fukazawa, Y., Hirayama, M., Honda, H., Ishisaki, Y., Kikuchi, K., Kubo, H., Murakami,  T., Ohashi, T., Takahashi, T. \& Yamashita, K., 1997, ApJ, 481, 660.
\bibitem[gir]{Gir} Girardi, M., Giurin, G., Mardirossian, F., Mezzetti, M. \& Boschin, W., ApJ, 1998, 505, 74.
\bibitem[Irwin et al. 1999]{irw} Irwin, J.A., Bregman, J.N. \& Evrard, A.E., 1999, ApJ in press, astro-ph/9901406.
\bibitem[Kemp \& Meaburn 1991]{kemp} Kemp, S.N. \& Meaburn, J., 1991, MNRAS, 251, 10. 
\bibitem[Kim \& Fabbiano 1995]{kim} Kim, D. \& Fabbiano, G., 1995, ApJ, 441, 182.
\bibitem[Markevitch et al. 1998]{tmaps} Markevitch, M., Forman, W., Sarazin, C. and Vikhlinin, A.,  1998, ApJ, 503, 77.
\bibitem[Markevitch \& Vikhlinin 1997]{A2256} Markevitch, M., \& Vikhlinin, A., 1997, ApJ, 491,467. 
\bibitem[Markevitch \& Vikhlinin 1997a]{mv1} Markevitch, M., \& Vikhlinin, A., 1997a, ApJ, 474, 84. 
\bibitem[Markevitch \& Vikhlinin 1997b]{A2256} Markevitch, M., \& Vikhlinin, A., 1997b, ApJ, 491,467. 
\bibitem[Markevitch et al. 1999]{A496_A2199} Markevitch, M.,  Vikhlinin, A., Forman, W. \& Sarazin, C.,  1999, ApJ, submitted, astro-ph/9904382.
\bibitem[Mathews 1978]{math} Mathews, W.G., 1978, ApJ, 219:413-423.
\bibitem[Mohr et al. 1999]{mohr} Mohr, J.J., Mathiesen, B., \& Evrard, A.E., 1999, ApJ, 517, 627.
\bibitem[Navarro et al. 1997]{nfw} Navarro, J., Frenk, C., White, S., 1997, ApJ, 490:493 (NFW). 
\bibitem[Neumann \& Arnaud 1999]{neu} Neumann, D.M. \& Arnaud, M., 1999, A\&A, 348, 711 . 
\bibitem[Nevalainen et al. 1999a]{A401} Nevalainen, J., Markevitch,M. \& Forman, W., 1999a, ApJ, in press, astro-ph/9906286.
\bibitem[Nevalainen et al. 1999b]{mtrel} Nevalainen, J., et al, 1999b, ApJ, in press, astro-ph/9911369.
\bibitem[Nevalainen \& Roos 1998]{nev} Nevalainen, J., \& Roos, M., 1998, A\&A, 339, 7.
\bibitem[Peres et al. 1998]{peres} Peres, C.B, Fabian, A.C., Edge, A.C., Allen, S.W., Johnstone, R.M. \& White, D.A., 1998, MNRAS, 298, 416.
\bibitem[Ponman \& Bertram 1993]{pon} Ponman, T.J. \& Bertram, D., 1993, Nature, 363, 51.
\bibitem[Raychaudhury et al. 1991]{ray} Raychaudhury, S., Fabian, A.C., Edge, A.C., Jones, C. \& Forman, W., 1991, MNRAS, 248, 101. 
\bibitem[Roos \& Harun-or-Rashid 1999]{matts} Roos, M., S. M. Harun-or-Rashid, 1999, International Europhysics Conference on High Energy Physics, Tampere, Finland, 15-21 July 1999, edited by K. Huitu, H.
Kurki-Suonio and J. Maalampi, IOP Publishing (Bristol, UK).
\bibitem[Quintana \& de Souza 1993]{quin} Quintana, H., \& de Souza, R, A\&ASS, 1993, 101, 475.
\bibitem[Sarazin 1988]{sar} Sarazin, 1988, {\it X-ray emissions from clusters of galaxies}, Cambridge University Press.
\bibitem[Sarazin, Wise \& Markevitch 1998]{sar2029} Sarazin, C.L., Wise, M.W. \& Markevitch, M.L., 1998, ApJ, 473, 651.
\bibitem[Snowden et al. 1994]{sno} Snowden, S.L., McCammon, D., Burrows, D.N. \& Mendenhall, J.,A, 1994, ApJ, 424:714-728.
\bibitem[Vikhlinin et al. 1999]{vikh} Vikhlinin, A., Forman, W. and Jones, C., 1999, ApJ, in press, astro-ph/9905200.
\bibitem[White et al.  1993]{wnef} White, S.D.M., Navarro, J.F., Evrard, A.E. \& Frenk, C.S., 1993, Nature 366, 429.
\bibitem[White \& Fabian 1995]{wf} White, D. and Fabian, A., 1995, MNRAS 273, 72.


\end{thebibliography}
\end{document}